\newcommand{\be}{\begin{equation}}
\newcommand{\ee}{\end{equation}}

\documentstyle[aps,epsf,multicol]{revtex}
\begin{document}
\draft
\widetext
\title{Localization, Coulomb interaction, topological principles and 
the quantum Hall effect.}
\author{A.M.M. Pruisken$^{1}$, M.A. Baranov$^{1,2}$ and I.S. Burmistrov$^{3}$ }

\address{$^{1}$ University of Amsterdam, Valckenierstraat 65, 1018XE Amsterdam, the Netherlands}
\address{$^{2}$ RRC "Kurchatov Institute", Kuchatov Square 1, 123182 Moscow, Russia}
\address{$^{3}$L.D. Landau Institute for Theoretical Physics, Kosygina str. 2, 117940 Moscow, Russia }

\maketitle

\begin{abstract}
\noindent{We report the consequences of a new interaction symmetry that 
protects the renormalization of the electron gas in low dimensions in general,
and in the quantum Hall regime in particular. We introduce a generalized 
Thouless' criterion for localization to include the effects of the Coulomb 
interaction and establish the quantization of the Hall conductance as well 
as the theory of massless edge bosons.}
\end{abstract}

\pacs{PACSnumbers 72.10.-d, 73.20.Dx, 73.40.Hm}

\begin{multicols}{2}
\narrowtext



Over the last few years, dramatic progress has been made\cite{EuroLett,paperI,paperII,paperIII} in 
theory of localization and interaction effects
in the quantum Hall regime. By now it is well understood that the
the Coulomb interaction problem falls in a new, non-Fermi liquid 
universality class of transport problems with a previously unrecognized
symmetry, called $\cal F$ invariance\cite{paperI}. Although superficially, the
results for scaling are very similar to those obtained for 
the free electron gas\cite{freepart}, 
it is important to bear in mind that we are dealing with two fundamentally
different problems and Fermi liquid principles are abscent. For example,
unlike the free particle problem, the field theory for the electron 
gas in the
presence of Coulomb interactions provides the foundations
for a unifying 
theory that includes both the
integral and fractional quantum Hall effect\cite{paperI,paperIII,fracedge}. 
The basic ingredients 
iof this theory are the Finkelstein approach to localization and
interaction phenomena\cite{Finkelstein}, the topological concept of an 
instanton vacuum\cite{freepart,instantons}
and the Chern Simons statistical gauge fields\cite{cf}.

In this Letter we present the detailed consequences of $\cal F$ invariance
and demonstrate the fundamental differences between the Coulomb 
interaction problem
and the free electron problem which effectively exist in different dimensions.
Moreover, we report new advances, obtained
from different sources, on the subject of renormalizability. These 
enable us to make contact with the theory
on the strong coupling side, notably the quantum Hall effect and the theory
of massless chiral edge bosons\cite{paperIII}.

In order to discuss the principle of $\cal F$ invariance, we consider the simpest case
of spin polarized or spinless electrons. Within the fermionic path integral representation we can
write the action as follows
$$      S[\bar{\Psi},\Psi,\lambda] =
        - \int_0^{1/T} d\tau \left[ \int_{x,x'} \lambda (\vec x ,\tau)
	U_0^{-1} (\vec x - \vec x') \lambda (\vec x ,\tau) \right.  
$$
\be     \left. +\int_x \bar{\Psi} [-\partial_\tau
        +i\lambda
	+\mu-{\cal H}(\vec x)
        -V(\vec x)] \Psi  \right]
	.
\ee

\noindent{Here,} $\tau$ stands for imaginary time, $\bar{\Psi} =\bar{\Psi} (\vec x ,\tau)$ and
$ {\Psi} = {\Psi} (\vec x ,\tau)$ are the fermion fields, $\mu$ is the chemical potential
and ${\cal H}(\vec x) + V(\vec x)$ denotes the free particle hamiltonian with a random impurity potential $V$. 
The electron-electron interactions are
represented by the field variable $\lambda = \lambda  (\vec x ,\tau)$  whereas $U_0^{-1}$ stands for 
the inverse of a interaction potential.

The action is invariant under the transformation

\be
\Psi \rightarrow e^{i\phi (\tau)} \Psi  \;\; ; \;\;\; \lambda \rightarrow \lambda + 
\partial_\tau  \phi(\tau) ,	
\ee
provided the Fourier transform of $U_0^{-1}$ vanishes at zero momentum, i.e. $U_0 ({\vec x}-{\vec x}')$ has
an infinite range. The invariance of the theory under global,
spatially independent gauge transformations has been named $\cal F$ {\em invariance}.

Following Finkelstein\cite{Finkelstein}, the effective quantum theory of metals is given in terms 
of unitary matrix field variables $Q_{n,m}^{\alpha\beta} (\vec x )$ replacing the composite variable
$\bar{\Psi}^\alpha (\vec x ,\omega_n ) {\Psi}^\beta (\vec x , \omega_m )$. Here, 
$\alpha, \beta$ represent the replica indices, commonly used to facilitate the averaging over random impurities; 
$n,m$ are the indices of the Matsubara frequencies
$\omega_{k} = \pi T(2k+1)$ with $k=n,m$. 

It is easy to understand how the manifold of the matrix $Q$ variables comes about. At a mean field level,
i.e. in the abscence of vertex corrections,
the metal is usually described by a ground state that breaks the $U(1)$ gauge invariance. 
It can be expressed as

\be
<Q_{n,m}^{\alpha\beta}>_{mf} = {\tau_0^{-1}} \Lambda_{n,m}^{\alpha\beta}
\ee
where $\tau_0$ stands for the elastic scattering time and the diagonal 
matrix $\Lambda_{n,m}^{\alpha\beta} = {\rm sign}(\omega_n )
\delta_{n,m} \delta_{\alpha\beta}$,
indicating that a branch cut has been generated separating the positive Matsubara frequencies from the
negative ones. 
Next, to obtain the lowest energy excitations of the metal, one may proceed by considering how 
$<Q>_{mf}$ transforms under $U(1)$ gauge transformations.
In Matsubara frequency space a $U(1)$ gauge
transformation can be represented by a 'large' unitairy matrix $W$

\be
W= exp (i \sum_{|n| < n_0 , \alpha} \phi^\alpha (\omega_n ) I_n^\alpha ) 
\ee
where the matrices $[ I_k^\gamma ]_{n,m}^{\alpha\beta} = \delta_{\alpha\gamma} \delta_{\beta\gamma}
\delta_{n,m+k}$ are the $U(1)$ generators. By 'large' we mean that the 'size' in frequency space 
($N_{max}$) of the matrices $W$ is taken to be arbitrarily large compared to the cut-off $n_0$
that is imposed on the gauge fields.
Write

\be
Q \rightarrow W^{-1} \Lambda W = t^{-1} U^{-1} \Lambda U t = t^{-1} \Lambda t.
\ee
Here, we have decomposed the $W$ into block diagonal or "longitudinal" matrices $U$ that only mixes
the frequecies in the $++$ and $--$ blocks, and "transverse" matrices $t$ that rotates 
amongst the positive and negetive frequencies. It follows that the 'size' ($n_{max}$) of the rotations $t$ 
can generally be taken to be much smaller than $N_{max}$, but much larger than the range of frequencies
$n_0$ that defines the $W$. The matrix field variables $Q= t^{-1} \Lambda t$
are the Goldstone modes of a broken $U(1)$ gauge invariance. 

Matrix manipulations in frequency space are generally non-trivial and gauge invariance demands that
the cut-offs $N_{max} >> n_{max} >> n_0$ are all taken to infinity in the end. We have developed a set of algebraic rules\cite{paperI}, named 
$\cal F$ algebra, that facilitate a general discussion of $\cal F$ invariance and $U(1)$ gauge invariance
in Matsubara frequency space. 

The quantum Hall effect is described by taking ${\cal H}(\vec x) 
= (\vec p - \frac{e}{c} \vec A (\vec x ))^2/2m $ in Eq. (1) with 
$\vec A$ representing a static magnetic field $B$. The
complete effective action of the $Q$ matrix fields, in the presence of 
external potentials $a_{\mu}^{\alpha}(\vec x , \omega_n)$, 
can then be written as

\be
	S_{eff} =  S_\sigma +S_F +S_U + S_0 .
\ee
Here, $S_{\sigma}$ stands for the ordinary non-linear sigma model action
\cite{freepart} 
\be
	S_{\sigma} = -\frac{\sigma_{xx}^0}{8} \!\int_x \!{\rm tr} [\vec D,Q]^2
	+\frac{\sigma_{xy}^0}{8} \!\int_x \! \epsilon_{ij} 
{\rm tr}  Q[D_i,Q][D_j,Q] ,	
\ee
where 
$D_j = \nabla_j -i \sum_{n\alpha} a_j^{\alpha} (\vec x , \omega_n ) I_n^{\alpha} $
is the covariant derivative.
Next, the two pieces $S_F$ and $S_U$  are interaction terms which are linear 
in the temperature $T$. 
The part $S_F$ is gauge invariant and
contains the singlet interaction term, first introduced by Finkelstein
\cite{Finkelstein}

\be
	S_{ F} = z \pi T 
	\int_x \left[ {\sum_{\alpha n}} {\rm tr} I_n^\alpha Q
	{\rm tr} I_{-n}^\alpha Q +4 {\rm tr}\eta Q - 6 {\rm tr}\eta \Lambda
	\right].
\ee
The piece $S_U$ contains both the interaction potential $U_0$ and the coupling to the scalar
potential
$$
	S_{U} = -\pi T {\sum_{n\alpha}}\int_{x,x'} U^{-1} (\vec x -\vec x' ) 
\times
$$
\be
	\left[
	{\rm tr} I_{-n}^\alpha Q(\vec x)-\frac{1}{\pi T} {\tilde a}_0^\alpha (\vec x , \omega_{-n} )
	\right]
	\left[
	{\rm tr} I_{n}^\alpha Q(\vec x')-\frac{1}{\pi T} {\tilde a}_0^\alpha (\vec x' , \omega_n )
	\right] .
\ee
The last part, $S_0$ is a magnetic field term
\be
	S_0 = -\frac{n_B^2 }{2\rho T}  
\int_x \sum_{\alpha n} b^{\alpha}(\vec x , \omega_n) 
	b^{\alpha}(\vec x , \omega_{-n}) .
\ee
We have defined (dropping the replica index $\alpha$ on $a_{\mu}$ from now 
onward)
\be
	\tilde{a}_0 =a_0 - i n_B  b / \rho \;\;\; ; \;\; U^{-1} (q) 
= (\rho^{-1} + U_0 (q))^{-1} .
\ee
Here, the density of states		
$\rho = {\partial n}/{\partial \mu}$ and the quantity $n_B = {\partial n}/{\partial B} $
are thermodynamic quantities.
The statement of gauge invariance now means that the theory is invariant under the following
transformation
\be
	Q \rightarrow W^{-1} Q W \;\;\; ; \;\;\; a_\mu \rightarrow a_\mu + \partial_\mu \phi .
\ee
Using this result, it is easy to see that  $\cal F$
invariance is broken if the interaction potential $U_0$ has a finite range, and 
retained otherwise.

The renormalizability of the theory can be established in different 
ways\cite{paperI,paperII}. 
The first approach, 
which is also the most physical one, is to study the response to the external 
potentials $S_{resp} (a_\mu)$, obtained after elimination of the Goldstone modes $Q$.
It is straightforeward to evaluate the theory at a tree level.
Defining the particle density $n$ as follows
$n = T { \delta S_{resp} }/{\delta  a_{0} }$,
then $n$ obeys the following continuity equation 
$$
	\omega_n ( n+i\sigma _{xy}^{0} b ) = 
$$
\be
      	{\vec \nabla} \cdot \left[ \sigma _{xx}^{0} ( {\vec  e} + {\vec \nabla} U_{0} n ) 
        - D_{xx}^0 {\vec \nabla} ( n + i n_B b ) \right] , 
\ee
with $D_{xx}^0 = \sigma_{xx}^0/\rho $ denoting the diffusion constant and 
${\vec  e}$, $b$ are the external electric and magnetic fields, 
respectively. 
This result is familiar from the 
theory of metals\cite{NozPines} where the quantity $n_B$ is usually neglected.
Notice that in the static limit $\omega_n \rightarrow 0$ both quantities $\sigma_{ij}^0$ 
drop out of the equation. The static response only contains the thermodynamic 
quantities $\rho, n_B$ and $U_0$ which are determined by an underlying theory that is completely
decoupled from the $Q$ field effective action. This means that the Coulomb interaction piece
$S_U$ does not have any quantum corrections in general, either perturbatively or
non-perturbatively. 
The only quantities that are allowed to have quantum corrections are the transport 
parameters $\sigma_{xx}^0$, $\sigma_{xy}^0$, and the singlet interaction amplitude $z$. 

As an important check on the statements of gauge invariance and renormalization,
we have evaluated the quantum theory in $2+\epsilon$ spatial dimensions to two loop order. 
The results of the computation, along with a complete analysis of dynamical scaling, will be 
reported elsewhere\cite{twoloop}. In what follows, however, we wish to address the subject of renormalizability
more formally, by making contact with the theory of ordinary non-linear sigma models. 
For this purpose we drop the external potentials from
the action and recall that for finite size matrices $Q$, operators like $S_F$ 
play the role of infrared regulators that
do not affect the singularity structure of the theory at short distances (i.e. the $\beta$ functions).
In particular, we know that the theory is renormalizable in two 
dimensions\cite{sigma}. 
Besides the renormalization of the coupling constant or $\sigma_{xx}^0$, 
one additional renormalization constant is needed for the operators 
linear in the $Q$ matrix field and two more renormalizations are generally needed for the operators 
bilinear in the $Q$ (i.e. the symmetric and anti-symmetric representation respectivily). 
These general statements formally apply to the theory with interactions as well 
since the latter only demands that the number of Matsubara frequencies is taken to infinity 
(along with the replica limit). To completely undust this point we have computed the 
cross over functions for the theory where the quantity $U^{-1} ({\vec x} - {\vec x}' )$ in $S_U$, Eq. (9),
is replaced by its most relevant part

\be
	U^{-1} ({\vec x} - {\vec x}' ) \rightarrow z \alpha \delta ({\vec x} - {\vec x}' ) .
\ee
Here,  $\alpha$ is an $\cal F$ invariance breaking parameter with $0 < \alpha < 1$ generally representing the
finite range interaction case. The extreme cases $\alpha = 0$ and $1$ describe the Coulomb system 
and the free electron gas respectivily.

In $2+\epsilon$ spatial dimensions we can write the renormalization group functions for the parameters $z$, $\alpha$
and the dimensionless resistance $g=\mu^\epsilon /\pi\sigma_{xx}$ (with $\mu$ an arbitrary momentum scale) 
as follows

$$
	\frac{dg}{d\ln\mu} = \epsilon g - g^2 \left(  f + f^2 + \frac{2 \alpha}{1- \alpha} 
	\ln (1-f +\alpha f)  \right) 
$$	
\be
	\frac{d\ln z}{d\ln\mu} = (1 - \alpha) g f,\;\;
	\frac{d \alpha}{d\ln\mu} = - (1- \alpha ) g (  1-f + \alpha f).
\ee
Here, $f = M^2/(\mu^2 + M^2)$ is a momentum scale dependent function with 
$M^2 = 8\pi z T n_{max}/\sigma_{xx}^0$. 
Notice that for $f=0$ ($\mu >> M$ or short distances) we obtain the well known
results for free particles\cite{freepart,instantons,sigma}: 
$dg/d\ln\mu$ has no one-loop contribution, 
$z$  has no quantum corrections in general and the result for $\alpha$ 
coincides with the renormalization of symmetric operators, bilinear in $Q$. 

On the other hand, for $f=1$ ($\mu << M$ or large distances), we obtain 
the peculiar Finkelstein 
results of the interacting electron gas\cite{paperII,italians}. 
The symmetry breaking parameter $\alpha$ now affects the renormalization 
of all the other parameters. The concept of $\cal F$ invariance 
($\alpha = 0$) now manifests itself as
a new (non-Fermi liquid) fixed point in the theory whereas the problem, 
for $0 < \alpha < 1$, 
lies in the domain of attraction
of the Fermi liquid line $\alpha =1$ which is stable in the infrared.

Notice, however, that the $\cal F$ invariant fixed point ($\alpha = 0$) 
exists only if the mass $M$ in the theory
remains finite at zero temperature. This clearly shows that, in order for 
$\cal F$ invariance
to represent an exact symmetry of the problem, $n_{max}$ must be infinite. 
The time $\tau$ therefore plays the role of an extra, non-trivial dimension 
and 
this dramatically complicates the problem of plateau
transitions in the quantum Hall regime. The Coulomb interaction problem is 
now given as a $2+1$ dimensional
field theory, thus invalidating any attempt toward exact solutions 
of the experimentally observed critical indices\cite{experiments}. 
This is in contrast to the theory with a finite range interaction potential 
($0 < \alpha < 1$)
which is truly two dimensional.

Next, we address the most interesting aspect of the theory, notably the 
$\sigma_{xy}^0$ term ($\theta$ term)
which was originally introduced in quantum Hall physics by Levine, Libby and 
one of us\cite{instantons}. The general problem
is quite analoguous to that of QCD where the topological concept of 
an {\em instanton vacuum} arose first. However,
in what follows, we shall make use of recent advancements in the theory of 
the quantum Hall effect\cite{largeN} and show that 
substantial progress can be made if it is recognized that the action, in 
the presence of the topological charge 

\be
C(Q) = \frac{i}{16\pi} \int_x \epsilon_{ij} {\rm tr} Q \nabla_i Q \nabla_j Q , 
\ee
generally distinguishes between {\em bulk} field configurations $Q_0$ that have an {\em integral} valued $C(Q_0 )$, 
and {\em edge} modes $q$ that carry an {\em unquantized} topological charge
$-\frac{1}{2} < C(q) < \frac{1}{2}$.
As was demonstrated in the context of the exactly solvable $CP^{N-1}$ theory with
large $N$\cite{largeN}, the bulk excitations $Q_0$ in the theory generally generate a massgap. On the other hand, the edge 
excitations $q$ always remain massless and they can be identified the familiar theory of {\em chiral edge bosons}\cite{paperIII,fracedge}. 
We shall now use these insights and unravel some of the outstanding strong coupling problems of our unifying theory. 
In particular, we are interested to see whether and how the theory {\em dynamically} displays the quantum Hall
state, along with a robust
quantization of the Hall conductance.

Following the large $N$ analysis\cite{largeN}, one proceeds by decomposing arbitrary field configurations $Q$ according to 
$Q=t^{-1} Q_0 t$ such that

\be
C(Q) = C(t^{-1} Q_0 t) = C(Q_0 ) + C( q ) ,
\ee
where $q=t^{-1} \Lambda t $. Along with the decomposition of $Q$, one recognizes also, from the microscopic
origins of the action, that the parameter $\sigma_{xy}^0$ is in fact given as the
sum of an exactly quantized {\em  edge} piece $m$ and an unquantized {\em bulk} piece $\theta$. 
It is helpful to think here of Landau level 
systems in a strong magnetic field inwhich case the 
parameter $\sigma_{xy}^0$ is identical to the filling fraction $\nu$. Write

\be
\sigma_{xy}^0 = \nu = m(\nu) +\frac{\theta(\nu)}{2\pi} ,
\ee
then $m(\nu)$ is an integer valued function and it equals $k$ if $\nu$ lies 
in the interval 
$k -\frac{1}{2} < \nu < k+ \frac{1}{2}$. The angle $\theta$ is confined to 
the interval 
$-\pi < \theta (\nu ) < \pi$. The $\sigma_{xy}^0$ term in the action, 
in the absence of a vector potential, can now be written as follows

\be
2\pi i \sigma_{xy}^0 C(Q) = i \theta (\nu) C(Q) + 2\pi i m(\nu) C(q).
\ee
Notice that the quantized {\em edge} piece $m(\nu)$ of $\sigma_{xy}^0$ is 
completely decoupled from the {\em bulk modes} $Q_0$ 
(it gives rise to trivial exponential phase factors that can be dropped). 
This quantity acts like a fixed point parameter of the theory whereas the 
angle 
$\theta(\nu)$ generally depends on the renormalization of the {\em bulk}
modes $Q_0$ in a non-perturbative manner (i.e. $\theta$ renormalization). 
In order to discuss this renormalization we first separate the {\em bulk} 
and {\em edge} pieces of the sigma model action and write

\be
	S_\sigma (Q) = {\tilde S}_{\sigma} (Q) + 2\pi i m(\nu) C(q) .
\ee
Here, ${\tilde S}_{\sigma}$ is the same as $S_\sigma$ except that 
$\sigma_{xy}^0$ is now being replaced by $\theta(\nu)/2\pi$. 

Next, in order to describe lowest energy excitations of the electron gas at 
zero temperature, we introduce an effective action for the edge modes $t$, 
obtained by formally eliminating the bulk field variables $Q_0$. Write

\be
	S_{eff} (q) = 2\pi i m(\nu) C(q) + S_{resp} (q), 
\ee
then

\be
e^{S_{resp} (q)} = \int_{\partial V} D[Q_0 ] e^{{\tilde S}_{\sigma} ( t^{-1} Q_0 t) + S_F (t^{-1} Q_0 t)} .
\ee
Here, the subscribt $\partial V$ on the integral sign indicates that the 
functional integral is being performed 
with a fixed value $Q_0 = \Lambda$ at the edge. It is important to remark 
that unlike $S_U$, one generally
can not leave out the singlet interaction term $S_F$ from this expression 
since it affects the renormalization of the theory at zero temperature. 

The result for the edge modes $t$ is precisely the same as the 
background field methodology, previously introduced for the purpose of 
defining the renormalization of the theory, both 
perturbatively\cite{paperII} and non-perturbatively (instantons)
\cite{EuroLett}. The action $S_{resp}$ has the same form as the original 
action,

\be
	{S_{resp} (q)} = {\tilde S}_{\sigma} ' (q) + S_F ' (q) ,
\ee
where the prime on the action indicates the "bare" parameters $\sigma_{xx}^0$, $\theta(\nu)$ and $z$ are now 
replaced by the "effective" ones $\sigma_{xx} '$, $\theta '$ and $z'$ respectivily.

Notice, however, that the "effective" theory truly describes the 
response of the system to a change in the 
boundary conditions. Since we expect the theory to dynamically generate 
a massgap for a range of values
about $\nu = integer$, we can introduce a generalized {\em Thouless 
criterion for localization} which says that
the $\cal F$ invariant system must be insensitive to changes in the boundary
conditions, i.e. $S_{resp}$ must be zero (or, rather, exponentially small 
in the system size). 
Under these circumstances, one can identify the quantity $m(\nu)$ 
in Eq. (21) as the quantized Hall conductance 
whereas $S_{eff} (q)$ is now recognized as the action for 
{\em massless edge excitations}\cite{paperIII}.

Notice that the criterion for localization is precisely consistent with the 
concept of $\theta$ renormalization\cite{EuroLett}, by instantons, which 
says that
at $\theta(\nu) = \pm \pi$, or $\nu =$ {\em half integer}, the system 
is quantum critical and a transition takes place between different 
quantum Hall states.

Several more steps are needed in order to make contact with the theory of 
{\em massless chiral edge bosons} $\phi_j$ 
that has been derived earlier\cite{paperIII,fracedge} 
under the assumption of an {\em energy gap}, rather than a
{\em mobility gap}. Here we just report the final result for the boson 
action, in the presence of external potentials ${\cal A}_\mu$,

$$
	S_{eff} ({\cal A}, \phi) = \frac{i}{4\pi T}  \int d\tau 
\biggl[ m(\nu) \int d^2 x \epsilon^{\mu\nu\kappa}
	{\cal A}_\mu \partial_\nu {\cal A}_\kappa 
$$
\be
-\oint dx \sum_{j=1}^{m(\nu)} (D_x \phi_j (D_\tau - i v_{eff} D_x )
	\phi_j - \phi_j {\cal E}_x ) 
\biggr],
\ee
where $D_{\mu}\phi_j=\partial_{\mu}\phi_j - {\cal A}_{\mu}$,
${\vec {\cal A}} = {\vec a}$, and ${\cal A}_0$ denotes an effective scalar 
potential ${\cal A}_0  = {\tilde a}_0  + (i m(\nu)/2 \pi) \int Ub$ with 
$v_{eff}$ representing an effective drift velocity that contains 
the interaction between the edge electrons.
The essence of this Letter lies in Eq. (24) which provides the complete 
answer to the fundamental questions regarding the precise quantization of 
the Hall conductance.

We are indebted to E. Brezin and A.M. Finkelstein for stimulating discussions. 
The research was supported in part by FOM and INTAS (Grant 99-1070).

\end{multicols}
\end{document}